\documentstyle[epsfig,natbib2,natbibmnfix]{mn2e}

\newcommand{\be}{\begin{equation}}
\newcommand{\ee}{\end{equation}}

\def\ltsima{$\; \buildrel < \over \sim \;$}
\def\simlt{\lower.5ex\hbox{\ltsima}}
\def\gtsima{$\; \buildrel > \over \sim \;$}
\def\simgt{\lower.5ex\hbox{\gtsima}}

\def\msun{{\,{\rm M}_\odot}}
\def\mbh{{\,{\rm M}_{\rm BH}}}

\def\del#1{{}}

\title{Clumpy winds and the obscuration of Active Galactic Nuclei.}

\author[S.~Nayakshin and J.~Cuadra]
{\parbox{18cm}{Sergei Nayakshin$^{1}$\footnotemark[1] and Jorge Cuadra$^{2}$}\vspace{0.3cm}\\
$^1$Department of Physics \& Astronomy, University of Leicester, Leicester, LE1 7RH, UK \\ 
$^2$Max-Planck-Institut f\"{u}r Astrophysik, Karl-Schwarzschild-Stra\ss{}e 1,
85741 Garching bei M\"{u}nchen, Germany} 

\begin{document}

\maketitle

\begin{abstract}

The role of star-formation driven outflows in the obscuration of the central
source in the Active Galactic Nuclei (AGN) is discussed. The outflow from a
sub-parsec scale accretion disc is numerically modelled for parameters
appropriate to the Galactic Centre. The resulting obscuration pattern is very
patchy, with some lines of sight becoming optically thick to Thomson
scattering. A fixed observer would see column depth changing by factors of
many over time scales of order months to hundreds of years, depending on
the physical size of the outflow region. Such winds may be relevant for
obscuration of some AGN and especially ``changing look AGN''.  However,
averaged over the sky as seen from the central source, these winds are always
optically thin unless wind outflow rates are super-Eddington. A simple
scaling argument shows that this is true not only for stellar-driven winds
but for any AGN winds. We therefore conclude that AGN winds are unable to
account for the vast majority of optically thick obscured AGN (a significant
fraction of all AGN). We suggest that the most likely source of optically
thick obscuration in AGN is a warped parsec scale accretion disc.
\end{abstract}

\renewcommand{\thefootnote}{\fnsymbol{footnote}}
\footnotetext[1]{E-mail: {\tt Sergei.Nayakshin at astro.le.ac.uk}}

\section{Introduction}
\label{intro}

According to the simplest version of the Unified model of Active Galactic
Nuclei (AGN), the central source has fundamentally the same properties in all
classes of
AGN \citep[e.g.,][]{Antonucci85,Antonucci93,Maiolino95,Risaliti99,Sazonov04}. It
is the orientation of our line of sight with respect to the dusty obscuring
environment of an AGN that makes it appear as a type I (relatively un-obscured)
or type II object (obscured). Despite of all the convincing observational
evidence for the existence of the absorber, most frequently thought to be
arranged in a shape of a torus, no convincing theoretical model has ever been
produced to explain its properties. Recently, \cite{Nenkova02} and
\cite{Risaliti02} demonstrated that to comply with the small observed torus sizes,
the absorber must not be uniform in density but should be rather clumpy,
i.e. consisting of individual smaller clouds. Theoretical models of clumpy tori
\citep[e.g.,][]{Krolik86,Krolik88,Vollmer04}, however, have all the same
fundamental difficulty. As the observations require the torus to be
geometrically thick, the velocity dispersion of the cold dusty clouds must be
of the order of their orbital velocity in the inner parsec of a galactic
centre. Optically thick tori would then have clouds colliding at super-sonic
speeds many times during one dynamical
time \citep{Krolik86,Krolik88}. Therefore, these models have to postulate
(somewhat arbitrarily, in our point of view), that the clouds can survive
these collisions, behaving in essence as rubber balls.

Recently, a number of authors argued that the obscuration must instead come
from clumpy winds \citep[e.g.,][]{Konigl94,Kartje99,Elvis00,Elvis04,Elitzur05}
emanating from the surface of an accretion disk, be these winds driven by line
pressure \citep{Proga03c}, continuum radiation pressure, heating, or by
hydromagnetic forces. In this paper we wish to critically assess this
suggestion. In particular, we consider AGN winds driven by star formation
activity in an accretion disc. Star or planet formation in AGN discs
has been long
predicted by theorists
\citep{Paczynski78,Kolykhalov80,Lin87,Collin99,Gammie01,Goodman03}, and recent
observations of the young massive stars in the centre of our Galaxy
\citep[e.g.,][]{Genzel03a,Ghez03b} lend a very strong support to these
theories \citep{Levin03,NS05,Paumard06}. Using the SPH treatment of
stellar winds developed by
\cite{Cuadra05,Cuadra06}, we calculate the obscuring properties of such winds for a
specific example. We find that these winds have very irregular, patchy
obscuration patterns. A fixed observer should then see the column depth
through the wind varying on time scales from months or years to tens and hundreds
of years, depending on the physical size of the wind launching region and the
SMBH mass.

However, observations (see \S 3 below) suggest that the structure responsible
for AGN obscuration should be optically thick over a large (e.g., a third)
fraction of the whole sky. Our models cannot achieve such a high average line
of sight column depth unless the mass loss rate in the wind is
super-Eddington. In fact, we show via simple analytical estimates that this
conclusion is valid whatever the nature of the wind driving mechanism is.  We
feel it would be very hard to produce highly super-Eddington winds in a
plausible quasi steady-state AGN model for a typical (very sub-Eddington
luminosity) AGN. Thus we suggest that winds, while very important for the AGN
phenomenon, cannot replace the ``torus'' in its role in the unification
schemes of AGN. Either the torus does exist despite the difficulties faced by
the models, or another optically thick structure, e.g., a warped accretion
disc plays its role in AGN.

\section{Stellar winds from AGN discs}\label{sec:method}

AGN discs are expected to be massive and thus self-gravitating at large enough
distances ($R \simgt 0.01-0.1$ pc) from the super-massive black holes
(SMBH). If the disc cooling time is short, gravitational collapse and
formation of stars or even planets is predicted
\citep[e.g.,][]{Paczynski78,Kolykhalov80,Shlosman89,Gammie01}. Young massive
stars in the central parsec of our own Galaxy were most likely created in this
way \citep{Paumard06}, with an apparent significant over-abundance of high
mass stars \citep{NS05} over their fraction in ``normal galactic'' star
formation. As stars are born from the accretion disc, they launch powerful
radiation fields and stellar winds, some of which will break through the
disc. Low mass proto-stars may prove equally effective in launching winds in
these circumstances as the rates at which gas is captured from the disk into
the Hill (or capture) zones of the protostars are super-Eddington
\citep{Nayakshin06a}. In addition, when the gas in the disk is depleted to
about half its initial surface density, stellar velocity dispersion starts to
grow, and soon the stellar disc becomes geometrically thicker than the gas
disc \citep{Nayakshin06a}. Stellar winds then escape from above the disc
directly.

We attempt to simulate this complex situation in a simplified setting,
concentrating only on the wind part of it. For this reason we do not include
the gas in the negligibly thin and flat accretion disc from which the stars
are born. To capture a degree of
the expected diversity of the stellar populations and conditions in this
problem, the stars are divided into two groups. The first produces winds with
velocities $v_{\rm w} = 300$ km/sec, whereas the second has $v_{\rm w} = 700$ km/sec. Both
types of stars have mass loss rates of $2.5 \times 10^{-4} \msun$ year$^{-1}$,
and are situated in a flat circularly rotating Keplerian disk of geometrical
thickness $H(R) = 0.1 R$ (the unit of length used here is 1 arcsecond at the
distance of 8 kpc, which is about $1.2\times 10^{17}$~cm or 0.04 pc). In
total, we have 200 mass shredding stars, thus amounting to the mass loss rate
of 0.1 $\msun$ year$^{-1}$, which is a factor of few super-Eddington for a
SMBH mass of $\mbh = 3.5 \times 10^6 \msun$.  The disc inner and outer radii
are $R_{\rm in} = 1.5$ and $R_{\rm out} = 8$, respectively. The stellar
surface density follows the law $\Sigma_*(R) \propto R^{-1}$, and the stars
are in the Keplerian circular rotation around the SMBH. The initial phase
(i.e. the $\phi$-coordinate) of the stars in the disk is generated randomly,
as is the vertical coordinate within the $-H$ to $+H$ limits. Initially, the
computational domain is filled with hot tenuous gas with velocity greater than
the escape velocity. This gas quickly outflows from the region. At the end of
the simulation stellar winds exceed the initial gas mass by a factor of few
tens.

To perform simulations, we use the Gadget-2 code \citep{Springel05} modified
to model stellar winds as new SPH particles ejected from the stellar wind
sources. These methods were described and thoroughly tested in
\cite{Cuadra05,Cuadra06}.

\begin{figure*}
\begin{minipage}[b]{.49\textwidth}
\centerline{\psfig{file=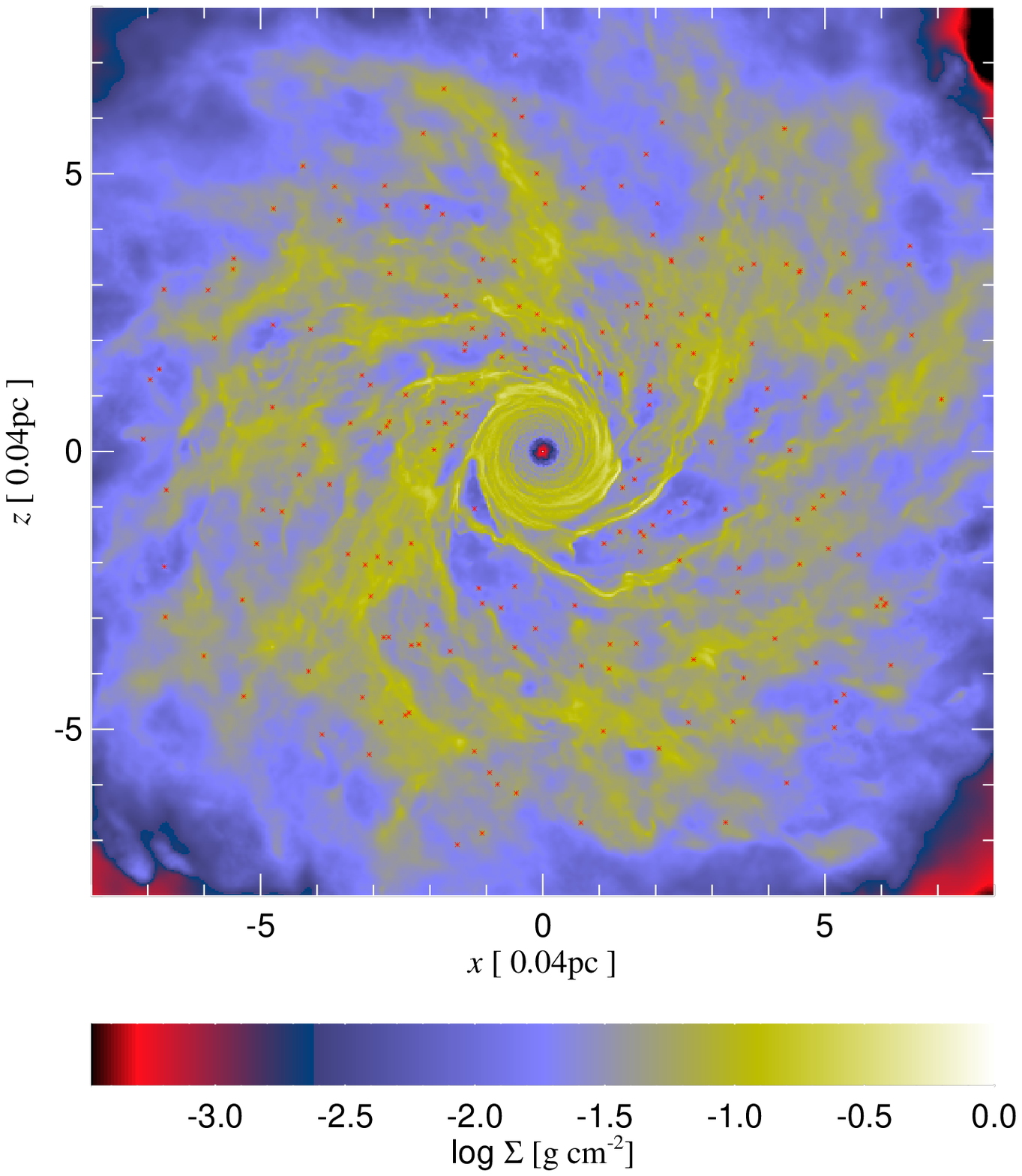,width=.99\textwidth,angle=0}}
\end{minipage}
\begin{minipage}[b]{.49\textwidth}
\centerline{\psfig{file=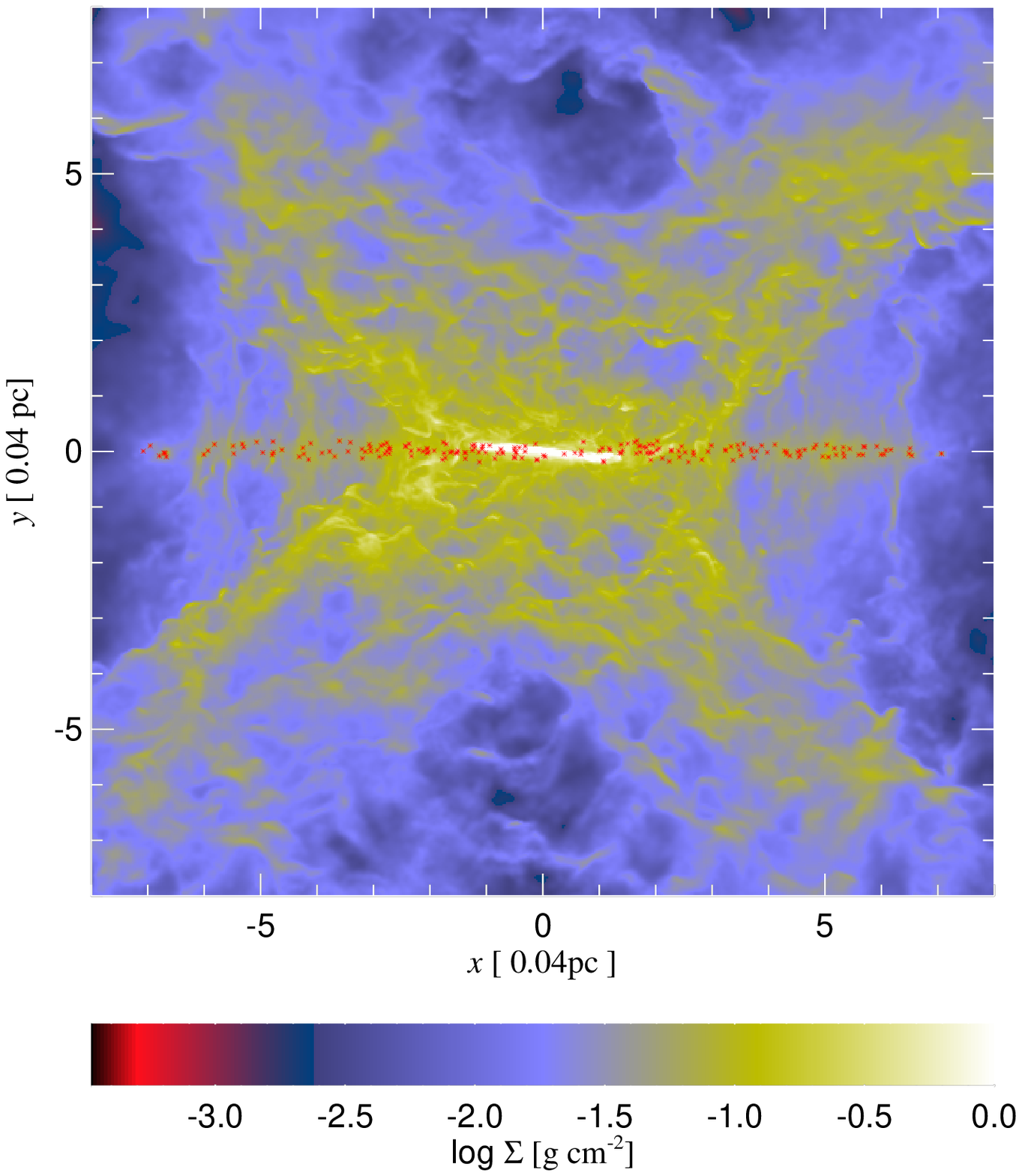,width=.99\textwidth,angle=0}}
\end{minipage}
\caption{Face-on (left) and side (right) views of the simulation domain for
  the flat stellar disc configuration. Red asterisks show the location of the
  wind-producing stars used in the simulation (the red ``dot'' in the centre
  of the left panel is not a star, but a low column depth region
  centred on the SMBH). }
\label{fig:fig1}
\end{figure*}

A snapshot of the simulation is shown in Figure \ref{fig:fig1} at time $t
\approx 2200$ years after the beginning of the simulation, which corresponds
to $\sim 36$ dynamical times at $R=1$. While there is no true steady state in
this finite number of moving stars system, the snapshot is fairly typical of
the morphology of the stellar wind. The face-on view (left panel) shows that
some of the shocked wind managed to cool down and formed a small-scale
disc. The inner part of the simulation domain is the most likely place for the
disc to form since stellar wind density is highest there, leading to shocks,
thermalization and rapid cooling; the frequent collisions of gaseous clumps
``waste'' momentum of the winds, and finally, the escape velocity from the
region is around 600 km/sec, i.e. higher than wind velocities of the slower
winds. With time the disc grows radially to both smaller and larger
radii. However, the formation of the disc \citep[cf.][]{Cuadra06} is
physically significant only in the case when the gaseous accretion disc from
which the stars were born is {\em entirely} absent, as is the case presently
in the Galactic Centre. In the opposite case the cooled disc and the ``spiral
arms'' seen at intermediate radii would simply blend in with the much more
massive underlying gas accretion disc.

Contrary to the frequent collisions and high escape velocity in the
inner region of the computational domain, the conditions in the outer regions
allow direct escape of both the faster diffuse and the slower cooler clumpy
winds.  Due to the final extent of the stellar disc and the projection effect,
the wind morphology reminds and ``X''-shape (Figure \ref{fig:fig1}, right
panel).  The edge-on view (right panel) of the stellar disc shows that the
plane of the gas disc at $R\simlt 2$ is very slightly tilted with respect to
the orbital plane of the stellar disc. This is an artefact of the initial
conditions.


\begin{figure*}
\begin{minipage}[b]{.49\textwidth}
\centerline{\psfig{file=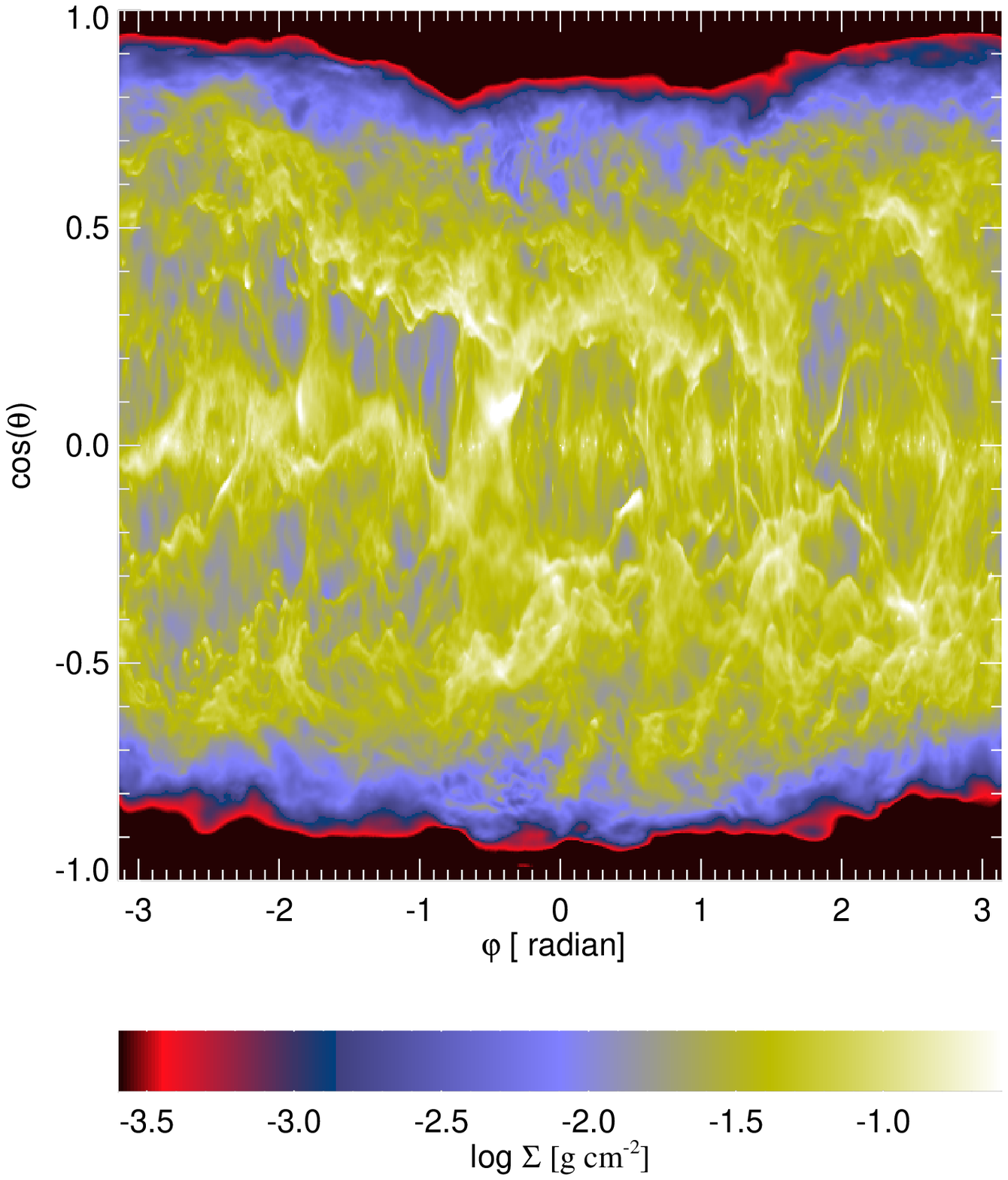,width=.99\textwidth,angle=0}}
\end{minipage}
\begin{minipage}[b]{.49\textwidth}
\centerline{\psfig{file=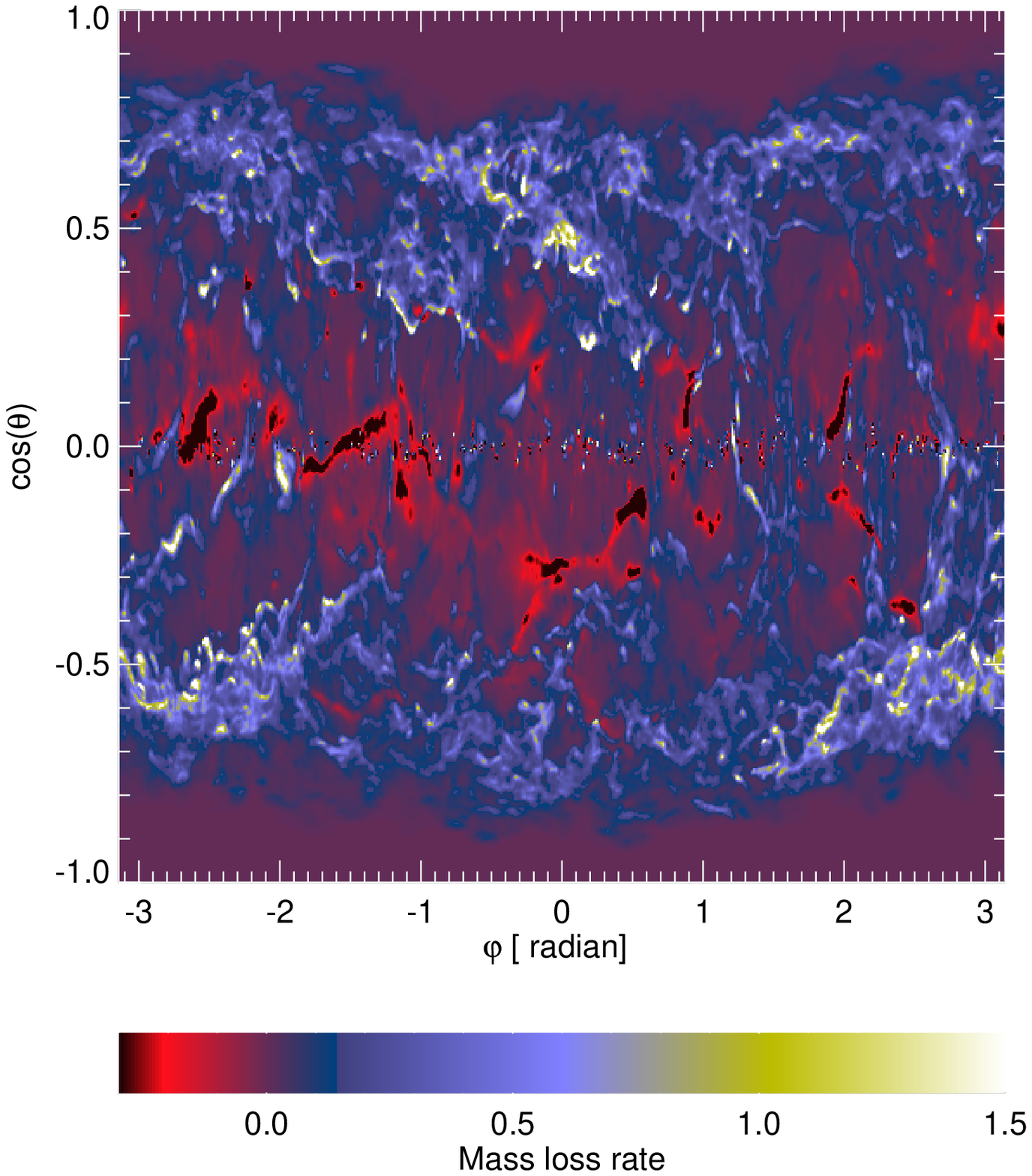,width=.99\textwidth,angle=0}}
\end{minipage}
\caption{Column depth through the wind (left) and isotropic mass outflow rate
  (right; in units of $\msun$ year$^{-1}$) for the simulation shown in
  Fig. \ref{fig:fig1}. Note the large variations of the obscuring column depth
  over small angular scales. The gaseous inner disc seen in the left panel of
  Figure \ref{fig:fig1} has been excluded from these Figures.}
\label{fig:fig2}
\end{figure*}

The left panel of Figure \ref{fig:fig2} shows the obscuring column depth of
the winds as seen from the SMBH. Since the orientation of the tilted inner disc is
influenced by the initial conditions, we eliminate its obscuration, 
only plotting gas with radial distances $R \simgt 1.6$ from the
SMBH, which excludes all of the disc.  Besides, had we included in the
simulations the more massive gas accretion disk whose midplane would coincide
with that of the stars, the inner disc seen in Figure \ref{fig:fig1} would
have completely blended in with the accretion disc and would not present much
obscuration at all.

Note the very irregular patchy structure of the (yellow) optically
thicker regions. The contrast between those and neighbouring less dense patches
of sky is frequently a factor of 10 or more. Since the winds are rotating at a
fraction of the local angular frequency, $\Omega$, this implies that the
column depth sampled by the observer will vary on time scales as short as
$\sim 10^{-3}$ to $10^{-2}$ of $1/\Omega$, i.e. $t_{\rm var} \sim 0.04-0.4\,
\hbox{year}\; (R/0.1\hbox{pc})^{3/2} M_8^{-1/2}$. The dotted pattern at
the $\cos \theta= 0$ plane are the dense regions of stellar winds immediately
next to the stars. 

The right panel of Figure \ref{fig:fig2} shows the isotropic mass loss rate
along the line of sight, defined as
\begin{equation}
\dot M \equiv \frac{\int d \Sigma \; 4 \pi R^2 \rho v_R}{\int d \Sigma}\;,
\label{defmdot}
\end{equation}
where the integral is taken along the line of sight defined by a given
$\theta$ and $\phi$. This quantity is useful for comparison with observations
of AGN outflows. An observer has only one line of sight available at any given
moment, and the assumption frequently made is that the outflow is roughly
isotropic, thus $\dot M \sim 4\pi R^2 \rho v_R$. We find that there are lines
of sight that yield isotropic mass outflow rates of $\sim 1-10 \msun$
year$^{-1}$, which is as much as a hundred times larger than the correct
sky-averaged value.  Also note that not all of the optically thicker regions
seen in the left panel of figure \ref{fig:fig2} appear equally prominently in
the outflow map in the right panel of the figure, as some of these structures
have a small radial velocity or are even infalling to smaller radii.

\section{Comparison to observations}\label{sec:analytics}

Recent multi-wavelength observations
\citep[e.g.,][]{Jaffe04,Packham05,Prieto05} indicate that AGN absorbers are
relatively small, parsec-scale structures, rather than the extended (100 to
1000 parsec) ones as was thought a decade earlier on the basis of observations
with poorer angular resolution.  \cite{Nenkova02} and \cite{Elitzur05}
demonstrate that this observational fact demands the ``torus'' to be clumpy in
order to allow high and low temperature regions to be simultaneously present
at the same location in the torus. \cite{Nenkova02} \citep[see
also][]{Elitzur05} show that the average number of clouds on the line of sight
to the nucleus, $N_{\rm a}$, should be of the order of a few to ten. Let $M_{\rm c}$ be the
mass of such a cloud, $M_{\rm c} = \Sigma_{\rm c} \pi R_{\rm c}^2$, and $\Sigma_{\rm c}$ and $R_{\rm c}$ be
the surface density and cloud radius, respectively. In the model of these
authors, the average column depth on the line of sight, $N_{\rm a} \Sigma_{\rm c}$ is of
the order of 1 g cm$^{-2}$ {\em or larger}, which is also reasonable given a large
fraction of AGN that are Thomson-thick
\citep[e.g.,][]{Sazonov04,Guainazzi05}. The total number of the clouds, $N_{\rm t}$,
is connected to $N_{\rm a}$ through
\begin{equation}
N_{\rm a} = \frac{N_{\rm t} \pi R_{\rm c}^2}{2\pi R_{\rm t}^2}\;,
\label{ntot}
\end{equation}
where $R_{\rm t}$ is the typical radius of the torus, and we assumed that it covers
roughly $2\pi$ of the sky as seen from the SMBH. Now, assume that the torus is
made up of clouds that form in the outflowing disk wind, as in the simulations
presented in this paper. The total mass outflow rate of the clouds is then
\begin{equation}
\dot M_{\rm wind} \sim 2\pi R_{\rm t}^2 N_{\rm a} \Sigma_{\rm c} \Omega_{\rm K} \sim 15
\frac{\msun}{\hbox{year}}\; N_{\rm a} \Sigma_{\rm c} \left(\frac{R_{\rm t}}{\hbox {1
pc}}\right)^{1/2} M_8^{1/2} \;,
\label{mwreq}
\end{equation}
where $M_8$ is the SMBH mass in units of $10^8 \msun$, and we assumed that
clouds radial velocities are of the order of the local Keplerian speed. Now, the
Eddington accretion rate is $4\pi G \mbh m_{\rm p}/\epsilon c \sigma_{\rm T} \approx
2 \msun \hbox{year}^{-1} M_8$. Thus the required wind mass loss rate (equation
\ref{mwreq}) is an order of magnitude higher, typically, than the Eddington
accretion rate for a $M_8 \sim 1$ object and a parsec-scale torus. The
estimate of $\dot M_{\rm wind}$ could be reduced somewhat by postulating even
smaller torus sizes, but it is hard to make $R_{\rm t}$ much smaller than $\sim 0.1$
pc as then the dust would be sublimated by the AGN radiation field
\citep[e.g., eq. 3.2 in][]{Emmering92}. However, we rather think that estimate
\ref{mwreq} is too optimistic, e.g. that an even higher mass outflow rates are
needed, as the observations and modelling of optically thick sources require a
{\em minimum} absorbing column depth, and hence in a good fraction of sources
$N_{\rm a} \Sigma_{\rm c}$ may actually be much higher than 1 g cm$^{-2}$.


Considering a specific case of the local obscured AGN studied by
\citep{Guainazzi05}, we note that the bolometric luminosities of these objects
in the infrared, X-ray and optical bands are in the range $L \sim 10^{43} -
\hbox {few} \times 10^{44}$ erg/sec, which implies SMBH accretion rates of
``only'' $ \sim 0.01 \msun$~year$^{-1}$ for the standard radiative
efficiency. Hence if the obscuration of the optically thick objects in that
sample were provided by the winds, we would conclude that the SMBH accretion
process must be very wasteful, with $\sim 100-10,000$ times more mass flowing
out of the inner parsec than accreting on the SMBH. \citep[Note that these
moderately bright AGN are not likely to be in the non-radiative accretion flow
regime when vigorous outflows are in fact expected, e.g., ][]{Blandford99} It
would also require a very high mass influx into the inner parsec to sustain
such winds. Given the difficulty of delivering enough fuel to the SMBHs even
in the earlier gas-rich epochs \citep{Thompson05}, it is hard to see how such
high mass influxes could be maintained.

\begin{figure}
\centerline{\psfig{file=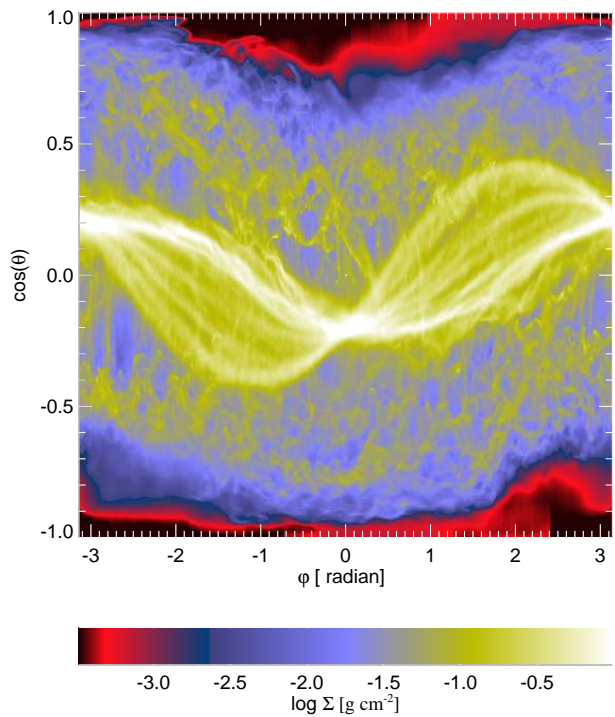,width=.5\textwidth,angle=0}}
\caption{Same as the left panel of Figure \ref{fig:fig2} but for slightly different initial
  conditions and with the cold tilted and warped inner disc included in the
  Figure. Notice that this very light (only $\sim 3 \msun$) disc is a very
  efficient absorber. }
\label{fig:fig3}
\end{figure}

\section{Discussion and Conclusions}\label{sec:conclusions}

Accreting black holes can drive strong outflows via X-ray heating, line or
continuous radiation pressure, and hydromagnetic forces
\citep[e.g.,][]{Begelman83,Konigl94,Kartje99,Proga03c}. Star formation on the
outskirts of cool massive accretion discs will also result in winds driven by
outflows from the young massive stars \citep{Cuadra05}. Here we studied the
obscuration properties of the AGN outflows of the last type. We found that
these outflows are quite clumpy, with the lines of sight passing through the
clumps becoming moderately optically thick to Thomson scattering for high
enough wind mass loss rates. Such outflows are bound to play a role in the
obscuration of AGN.

We however feel that this role cannot be dominant as the sky averaged column
depth is significantly smaller than the one required by the current ``torus''
models \citep[e.g.,][]{Nenkova02,Elitzur05} and observations
\citep{Guainazzi05} for realistic mass outflow rates, e.g. comparable to the
accretion rates onto the SMBH, which are likely to be only $0.01-0.1$ of the
Eddington accretion rates in the local AGN \citep[and are much smaller if we
also consider Low Luminosity AGN; see, for example, Fig. 1.9 in ][]{Ho05}. We
thus believe that the search for the culprit of the optically thick
obscuration of AGN should still go on.

One rather obvious candidate is a warped accretion disc. The disc may be
significantly warped by instabilities due to back reaction to mass outflow or
due to the AGN radiation pressure, or it may be warped due to precession in a
non-axisymmetric gravitational potential
\citep[e.g.][]{Schandl94,Pringle96,Nayakshin05}. Finally, there is no reason
for an initial disc configuration to be flat if the mass deposition is
much shorter than the disc viscous time, which was probably the case in the
Galactic Centre \citep{NC05,Paumard06}. A simulation ran by us for the present
paper and later rejected (because the initial mass of the gas turned out too
high so that it influenced the outcome of the simulation) is curiously useful
in this respect. Figure \ref{fig:fig3} shows the obscuration pattern of this
simulation in the same way as in the left panel in Figure \ref{fig:fig2}, but
including the tilted and slightly warped gas disc. Except for the initial
conditions, this simulation is exactly identical to the one shown in Figures 1
\& 2. As Figure \ref{fig:fig3} shows, the obscuration provided by the winds
literally pales in comparison with that produced by the warped disc, whose
mass is only $\sim 3 \msun$. An accretion disc is a long lived structure that
can accumulate a lot of gas (still a small fraction of the SMBH mass, however)
and can be Thomson-thick out to parsec distances. Obscuration by a warped
accretion disc requires {\em no} continuous energy input, and would work for
low or high luminosity AGN as long as the discs are sufficiently massive and
strongly warped.

The authors acknowledge discussions and comments from Ken Pounds and Sergei
Sazonov.

\bibliographystyle{mnras}
\bibliography{../nayakshin}

\begin{thebibliography}{45}
\expandafter\ifx\csname natexlab\endcsname\relax\def\natexlab#1{#1}\fi

\bibitem[{Antonucci}(1993)]{Antonucci93}
{Antonucci} R., 1993, \araa, 31, 473

\bibitem[{Antonucci} \& {Miller}(1985)]{Antonucci85}
{Antonucci} R.~R.~J., {Miller} J.~S., 1985, \apj, 297, 621

\bibitem[{Begelman} et~al.(1983){Begelman}, {McKee} \& {Shields}]{Begelman83}
{Begelman} M.~C., {McKee} C.~F., {Shields} G.~A., 1983, \apj, 271, 70

\bibitem[{Blandford} \& {Begelman}(1999)]{Blandford99}
{Blandford} R.~D., {Begelman} M.~C., 1999, \mnras, 303, L1

\bibitem[{Collin} \& {Zahn}(1999)]{Collin99}
{Collin} S., {Zahn} J., 1999, A\&A, 344, 433

\bibitem[{Cuadra} et~al.(2005){Cuadra}, {Nayakshin}, {Springel} \& {Di
  Matteo}]{Cuadra05}
{Cuadra} J., {Nayakshin} S., {Springel} V., {Di Matteo} T., 2005, \mnras, 360,
  L55

\bibitem[{Cuadra} et~al.(2006){Cuadra}, {Nayakshin}, {Springel} \& {di
  Matteo}]{Cuadra06}
{Cuadra} J., {Nayakshin} S., {Springel} V., {di Matteo} T., 2006, \mnras, 366,
  358

\bibitem[Elitzur(2005)]{Elitzur05}
Elitzur M., 2005, astro-ph/0512025

\bibitem[{Elvis}(2000)]{Elvis00}
{Elvis} M., 2000, \apj, 545, 63

\bibitem[{Elvis} et~al.(2004){Elvis}, {Risaliti}, {Nicastro}, {Miller}, {Fiore}
  \& {Puccetti}]{Elvis04}
{Elvis} M., {Risaliti} G., {Nicastro} F., {Miller} J.~M., {Fiore} F.,
  {Puccetti} S., 2004, \apjl, 615, L25

\bibitem[{Emmering} et~al.(1992){Emmering}, {Blandford} \&
  {Shlosman}]{Emmering92}
{Emmering} R.~T., {Blandford} R.~D., {Shlosman} I., 1992, \apj, 385, 460

\bibitem[{Gammie}(2001)]{Gammie01}
{Gammie} C.~F., 2001, \apj, 553, 174

\bibitem[{Genzel} et~al.(2003){Genzel}, {Sch{\" o}del}, {Ott}
  et~al.]{Genzel03a}
{Genzel} R., {Sch{\" o}del} R., {Ott} T., et~al., 2003, \apj, 594, 812

\bibitem[{Ghez} et~al.(2003){Ghez}, {Duch{\^ e}ne}, {Matthews} et~al.]{Ghez03b}
{Ghez} A.~M., {Duch{\^ e}ne} G., {Matthews} K., et~al., 2003, \apj, 586, L127

\bibitem[{Goodman}(2003)]{Goodman03}
{Goodman} J., 2003, \mnras, 339, 937

\bibitem[{Guainazzi} et~al.(2005){Guainazzi}, {Matt} \& {Perola}]{Guainazzi05}
{Guainazzi} M., {Matt} G., {Perola} G.~C., 2005, \aap, 444, 119

\bibitem[{Ho}(2005)]{Ho05}
{Ho} L.~C., 2005, Ap\&SS, 300, 219

\bibitem[{Jaffe} et~al.(2004){Jaffe}, {Meisenheimer}, {R{\"o}ttgering}
  et~al.]{Jaffe04}
{Jaffe} W., {Meisenheimer} K., {R{\"o}ttgering} H.~J.~A., et~al., 2004, \nat,
  429, 47

\bibitem[{Kartje} et~al.(1999){Kartje}, {K{\"o}nigl} \& {Elitzur}]{Kartje99}
{Kartje} J.~F., {K{\"o}nigl} A., {Elitzur} M., 1999, \apj, 513, 180

\bibitem[{Kolykhalov} \& {Sunyaev}(1980)]{Kolykhalov80}
{Kolykhalov} P.~I., {Sunyaev} R.~A., 1980, Soviet Astron. Lett., 6, 357

\bibitem[{K\"onigl} \& {Kartje}(1994)]{Konigl94}
{K\"onigl} A., {Kartje} J.~F., 1994, \apj, 434, 446

\bibitem[{Krolik} \& {Begelman}(1986)]{Krolik86}
{Krolik} J.~H., {Begelman} M.~C., 1986, \apjl, 308, L55

\bibitem[{Krolik} \& {Begelman}(1988)]{Krolik88}
{Krolik} J.~H., {Begelman} M.~C., 1988, \apj, 329, 702

\bibitem[{Levin} \& {Beloborodov}(2003)]{Levin03}
{Levin} Y., {Beloborodov} A.~M., 2003, \apj, 590, L33

\bibitem[{Lin} \& {Pringle}(1987)]{Lin87}
{Lin} D.~N.~C., {Pringle} J.~E., 1987, \mnras, 225, 607

\bibitem[{Maiolino} \& {Rieke}(1995)]{Maiolino95}
{Maiolino} R., {Rieke} G.~H., 1995, \apj, 454, 95

\bibitem[{Nayakshin}(2005)]{Nayakshin05}
{Nayakshin} S., 2005, \mnras, 359, 545

\bibitem[{Nayakshin}(2006)]{Nayakshin06a}
{Nayakshin} S., 2006, ArXiv Astrophysics e-prints

\bibitem[{Nayakshin} \& {Cuadra}(2005)]{NC05}
{Nayakshin} S., {Cuadra} J., 2005, \aap, 437, 437

\bibitem[{Nayakshin} \& {Sunyaev}(2005)]{NS05}
{Nayakshin} S., {Sunyaev} R., 2005, \mnras, 364, L23

\bibitem[{Nenkova} et~al.(2002){Nenkova}, {Ivezi{\'c}} \& {Elitzur}]{Nenkova02}
{Nenkova} M., {Ivezi{\'c}} {\v Z}., {Elitzur} M., 2002, \apjl, 570, L9

\bibitem[{Packham} et~al.(2005){Packham}, {Radomski}, {Roche}
  et~al.]{Packham05}
{Packham} C., {Radomski} J.~T., {Roche} P.~F., et~al., 2005, \apjl, 618, L17

\bibitem[{Paczy\'nski}(1978)]{Paczynski78}
{Paczy\'nski} B., 1978, Acta Astron., 28, 91

\bibitem[{Paumard} et~al.(2006){Paumard}, {Genzel}, {Martins}, {Nayakshin} \&
  {et al}]{Paumard06}
{Paumard} T., {Genzel} R., {Martins} F., {Nayakshin} S., {et al}, 2006,
  submitted to ApJ (astro-ph/0601268)

\bibitem[{Prieto} et~al.(2005){Prieto}, {Maciejewski} \& {Reunanen}]{Prieto05}
{Prieto} M.~A., {Maciejewski} W., {Reunanen} J., 2005, \aj, 130, 1472

\bibitem[{Pringle}(1996)]{Pringle96}
{Pringle} J.~E., 1996, \mnras, 281, 357

\bibitem[{Proga}(2003)]{Proga03c}
{Proga} D., 2003, \apj, 585, 406

\bibitem[{Risaliti} et~al.(2002){Risaliti}, {Elvis} \& {Nicastro}]{Risaliti02}
{Risaliti} G., {Elvis} M., {Nicastro} F., 2002, \apj, 571, 234

\bibitem[{Risaliti} et~al.(1999){Risaliti}, {Maiolino} \&
  {Salvati}]{Risaliti99}
{Risaliti} G., {Maiolino} R., {Salvati} M., 1999, \apj, 522, 157

\bibitem[{Sazonov} \& {Revnivtsev}(2004)]{Sazonov04}
{Sazonov} S.~Y., {Revnivtsev} M.~G., 2004, \aap, 423, 469

\bibitem[{Schandl} \& {Meyer}(1994)]{Schandl94}
{Schandl} S., {Meyer} F., 1994, \aap, 289, 149

\bibitem[{Shlosman} \& {Begelman}(1989)]{Shlosman89}
{Shlosman} I., {Begelman} M.~C., 1989, \apj, 341, 685

\bibitem[{Springel}(2005)]{Springel05}
{Springel} V., 2005, \mnras, 364, 1105

\bibitem[{Thompson} et~al.(2005){Thompson}, {Quataert} \& {Murray}]{Thompson05}
{Thompson} T.~A., {Quataert} E., {Murray} N., 2005, \apj, 630, 167

\bibitem[{Vollmer} et~al.(2004){Vollmer}, {Beckert} \& {Duschl}]{Vollmer04}
{Vollmer} B., {Beckert} T., {Duschl} W.~J., 2004, \aap, 413, 949

\end{thebibliography}

\end{document}